\documentstyle[12pt,epsf]{article}
\begin{document}
\pagestyle{empty}
\begin{center}
\hspace {10.cm} DFUPG 107-95\\
\hspace {10.cm} UBC/GS/122-95\\
\hspace {10.cm} hep-th/9512048
\vspace {0.5 cm}

{\bf Charge Screening and Confinement in Hot 3-D QED }

\vspace {1.7 cm}
{\large G. Grignani\footnote{grignani@perugia.infn.it}$^a$,
G. Semenoff\footnote{semenoff@physics.ubc.ca}$^b$,
P. Sodano\footnote{sodano@perugia.infn.it}$^a$ and
O. Tirkkonen\footnote{tirk@physics.ubc.ca}$^b$}\\
\vspace {0.4cm}
$(a)$Dipartimento di Fisica and Sezione I.N.F.N.\\Universit\`a di
Perugia\\ Via A. Pascoli I-06123, Perugia, Italia\\~~\\ $(b)$Department
of Physics, University of British Columbia\\Vancouver, British
Columbia, Canada V6T 1Z1 \\
\vspace {0.4cm}
\end{center}
\vspace {0.4cm}
\centerline{\bf Abstract}

\vspace {0.5cm}

We examine the possibility of a confinement-deconfinement phase
transition at finite temperature in both parity invariant and
topologically massive three-dimensional quantum electrodynamics.  We
review an argument showing that the Abelian version of the Polyakov
loop operator is an order parameter for confinement, even in the
presence of dynamical electrons.  We show that, in the parity
invariant case, where the tree-level Coulomb potential is logarithmic,
there is a Berezinskii-Kosterlitz-Thouless transition at a critical
temperature ($T_c=e^2/8\pi+{\cal O}(e^4/m)$, when the ratio of the
electromagnetic coupling and the temperature to the electron mass is
small). Above $T_c$ the electric charge is not confined and the system
is in a Debye plasma phase, whereas below $T_c$ the electric charges
are confined by a logarithmic Coulomb potential, qualitatively
described by the tree-level interaction.  When there is a topological
mass, no matter how small, in a strict sense the theory is not
confining at any temperature; the model exhibits a screening phase,
analogous to that found in the Schwinger model and two-dimensional QCD
with massless adjoint matter. However, if the topological mass is much
smaller than the other dimensional parameters, there is a temperature
for which the range of the Coulomb interaction changes from the
inverse topological mass to the inverse electron mass.  We speculate
that this is a vestige of the BKT transition of the parity-invariant
system, separating regions with screening and deconfining
behavior.

\vspace{1.5 cm}

\centerline{PACS: 11.10.Wx; 11.15.-q}
\newpage
\pagestyle{plain}

\section{Introduction}

One of the most intriguing features of a gauge theory is the
possibility of confinement.  In a confining system, there are no
``in'' or ``out'' fields appearing in the asymptotic states which have
color charges ~\cite{conf,degi}.  As a result, all asymptotic states are
singlets under the symmetry transformations in the color group.  This
generally occurs in one of two ways.  First, as it is widely believed
to be the case in four-dimensional quantum chromodynamics (QCD), the
charged fields appearing in the bare action - quarks and gluons - are
permanently confined into color singlet bound states - mesons and
baryons - which make up the entire spectrum of asymptotic states.
Second, as well aso the appearance of color singlet bound states, it
is possible that the charges of bare fields are completely screened,
so that they interpolate physical fields which occur in the spectrum
but create color singlet states.  This possibility has been raised for
the electron field in the Schwinger model ~\cite{abd} and in two
dimensional adjoint QCD where the bare quark mass is zero \cite{gkms}.
This situation is often referred to as ``screening'' rather than
confinement.

At finite temperature, the difference between a confined and
deconfined phase is less evident than at zero temperature.  There is
no concept of asymptotic states and the quantitative observable
features are thermodynamic variables and correlation functions
governing the propagation of external influences.  The commonly used
test for confinement in a gauge theory at finite temperature is its
ability to screen static external charges.  The operator which probes
electric screening is the Polyakov loop operator ~\cite{po,su} which
is the trace of the path ordered exponential of the gauge field on a
path which links the periodic Euclidean time~\footnote{We use units
where Planck's constant, the speed of light and Boltzmann's constant
are one.  For a discussion of the path integral formulation of finite
temperature gauge theory see ~\cite{gpy}.}
\begin{equation}
P(\vec x)={\rm tr}{\cal P}{\rm ~e}^{i\int_0^{1/T}d\tau A_0(\tau,\vec x)}
\end{equation}
The expectation value of this operator is the exponential of the free
energy, $F(\vec x)$, which is required to immerse a classical
fundamental representation quark source in the gauge field medium at
the point $\vec x$,
\begin{equation}
\left\langle P(\vec x)\right\rangle = {\rm ~e}^{-F(\vec x)/T}
\end{equation}
If the expectation value is zero, corresponding to infinite free
energy, this is interpreted as a signal of confinement.  In this case,
the system is not capable of screening the electric flux which is
necessarily created with the external quark source and the electric
flux takes up a configuration which has infinite energy.  It is
important that the infinite energy arises from an infrared, rather
than ultraviolet divergence since the latter occurs even in
non-confining theories and could be cured by introducing a fundamental
cutoff.

If the expectation value of the Polyakov loop is non-zero, and the
free energy finite, this is interpreted as the system being in a
deconfined phase.  The electric flux associated with the source is
screened by the medium.

In compact gauge theories, the Polyakov loop operator can be used as
an order parameter for confinement in either pure Yang-Mills
gluo-dynamics or in a matter-coupled gauge theory when the matter is
in either the adjoint representation, or some other representation
whose degrees of freedom transform trivially under the center of the
gauge group.  In these cases, there is an invariance of the finite
temperature path integral under gauge transformations which twist by
an element of the center of the gauge group in the periodic Euclidean
time.  The operators in the action are all invariant and remain
periodic (or anti-periodic in the case of fermions) under such a
twisted gauge transform, but the operator $P(\vec x)$ is transformed
to $ZP(\vec x)$ where $Z$ is the central element.  Thus, the finite
temperature theory has an effective global symmetry whose
transformations are the cosets of the set of all gauge transformations
under which the path integral is symmetric modulo those which are
strictly periodic in Euclidean time.  Whether this symmetry is broken
or not is a well defined question.  Spontaneous breaking of the
symmetry is related to deconfinement ~\cite{po,su} and $P(\vec x)$ is
an order parameter.  This order parameter has been particularly useful
in characterizing the nature ~\cite{ftr}-\cite{sve} and quantitative
aspects ~\cite{weiss}-\cite{dadda2} of the confinement-deconfinement
phase transition in a wide array of pure gauge theories.

When the gauge theory is coupled to matter fields which transform
non-trivially under the center of the group, the symmetry is broken
explicitly and its realization can no longer be used as a probe for
confinement.  An example is QCD with quarks in the fundamental
representation of SU(3).  In that case, the question of distinguishing
a confining and non-confining phase of the finite temperature gauge
theory is more sophisticated ~\cite{nw}.

Recently it has been argued \cite{sm,gss} that the Abelian analog
of the Polyakov loop operator,
\begin{equation}
P_{\tilde e}(\vec x)\equiv {\rm ~e}^{i {\tilde e}\int_0^{1/T}d\tau
~A_0(\tau,\vec x)}
\label{pol}
\end{equation}
could be used as an order parameter for confinement in Abelian gauge
theories, even in the presence of dynamical charged particles. The
requirements are that the dynamical charged fields in the gauge
theory must have charges which are integral multiples of some basic
charge, which we denote by $e$.  Also, a technical requirement is that
the charged fields have a mass gap and that the field theory has a
finite ultraviolet cutoff.  In the limit where the cutoff is removed,
it is usually necessary to define the loop operator by multiplicative
renormalization.  In order that (\ref{pol}) be an order parameter, the
charge $\tilde e$ appearing there {\bf must not be an integer multiple
of the basic charge}: $\tilde e \neq e\cdot{\rm integer}$.

The expectation value of the operator (\ref{pol}) measures the
response of an electrodynamic system to placing a classical {\it
incommensurate} charge $\tilde e$ at point $\vec x$.  The quantity
\begin{equation}
F_{\tilde e}(\vec
x)=-T\ln \left\langle P_{\tilde e}(\vec x)\right\rangle
\end{equation}
is the free energy of the system in the presence of the classical
charge (minus the free energy when the charge is absent).  If this
free energy is finite, the system is not confining.  If the free
energy is infinite, this implies that the expectation value
(\ref{pol}) must be zero.  This means that it takes an infinite amount
of energy to immerse a classical charge in the system, implying that
it is in a confining phase.  If the charge in the loop operator were a
commensurate, rather than incommensurate one, its electric field could
be screened by producing a finite number of dynamical charged
particle-antiparticle pairs, using the appropriate number of particles
or antiparticles to screen the external charge and allowing the
remaining dynamical particles to escape to infinity.  This process
would take a finite amount of energy.  For this reason, we expect that
the loop operator with a commensurate charge would always have a
non-zero expectation value.  At zero temperature, the pair production
would take a threshold energy of the mass of the particles produced.
At finite temperature the thermally activated particles are already
present so that screening of this sort takes a small amount of energy.

Further information can be obtained from the correlators,
\begin{equation}
F_{\tilde e_1\dots \tilde e_n}(\vec x_1,\ldots,\vec x_n)
=-T\ln\left\langle \prod_{i=1}^n P_{\tilde
e_i}(\vec x_i)\right\rangle
\end{equation}
which give the electrostatic energy of an array of charges $\tilde
e_i$ situated at points $\vec x_i$, respectively.  For example, the
two-point correlator gives the effective potential between a positive
and negative charge,
\begin{equation}
V_{\tilde e,-\tilde e}(\vec x,\vec y)\equiv F_{\tilde e,-\tilde
e}(\vec x,\vec y)=-T\ln
\left\langle P_{\tilde e}(\vec x) P_{-\tilde e}(\vec y)\right\rangle
\end{equation}
If the clustering property of this expectation value holds, vanishing
or non-vanishing of the expectation value of a single loop operator is
related to the asymptotic behaviour of the potential: if
\begin{equation}
\lim_{\vert \vec x-\vec
y\vert\rightarrow\infty}V_{\tilde e,-\tilde e} (\vec x,\vec y) =\infty
\end{equation}
the two point function has the behavior
\begin{equation}
\lim_{\vert x-y\vert\rightarrow\infty}~
\langle P_{\tilde e}(\vec x) P_{-\tilde e}(\vec y) \rangle=0
\end{equation}
and the cluster decomposition property implies that the expectation
value of the single operator should vanish
\begin{equation}
\langle P_{\tilde e}(\vec x)\rangle~=~0
\end{equation}  
This characterizes confinement.  If, on the other hand
\begin{equation}
\lim_{\vert \vec x-\vec
y\vert\rightarrow\infty}V_{\tilde e,-\tilde e} (\vec x,\vec y)=~{\rm
finite~constant}
\end{equation}
the potential is not confining, the two-point correlator has the 
behavior
\begin{equation}
\lim_{\vert x-y\vert\rightarrow\infty}~
\langle P_{\tilde e}(\vec x) P_{-\tilde e}(\vec y) 
\rangle~=~{\rm constant}\ \ ,
\end{equation}
clustering implies that
\begin{equation}
\langle P_{\tilde e}(\vec x)\rangle~=~{\rm constant}
\end{equation}  
and an isolated external incommensurate charge has finite energy.

The expectation value of the loop operator in (\ref{pol}) is governed
by a particular discrete global symmetry, $Z$, isomorphic to the
additive group of the integers, which appears in the euclidean path
integral at finite temperature.  To see its origin, consider the
euclidean finite temperature functional integral expression for the
partition function ~\cite{gpy},
\begin{equation} Z[T]=\int \prod_{\vec
x,\tau\in[0,1/T]}[dA_\mu(\tau,\vec x) d\psi(\tau,\vec x)
d\bar\psi(\tau,\vec x)] {\rm ~e}^{-S[A,\psi\bar\psi]} \label{pi}
\end{equation}
where the action is
\begin{equation}
S[A,\psi,\bar\psi]=\int_0^{1/T}d\tau \int d\vec
x\left(\frac{1}{4}F_{\mu\nu}^2 (\tau,\vec x) +\bar\psi(\tau,\vec
x)\left(\gamma_\mu D_\mu+m\right)\psi(\tau,\vec x) \right)
\end{equation}
with $D_\mu=\partial_\mu-ieA_\mu$ and $F_{\mu\nu}=\nabla_\mu
A_\nu-\nabla_\nu A_\mu$.  The boundary conditions in time are periodic
for the photon,
\begin{equation}
A_\mu(\tau=1/T,\vec x)=A_{\mu}(0,\vec x)
\end{equation}
and antiperiodic for the electron,
\begin{equation}
\psi(\tau=1/T,\vec
x)=-\psi(\tau=0,\vec x) ~~,~~ \bar\psi(\tau=1/T,\vec
x)=-\bar\psi(\tau=0,\vec x)
\end{equation}

The path integral (\ref{pi}) is symmetric under gauge transformations,
\begin{equation}
A_\mu '(\tau,\vec x)=A_\mu(\tau,\vec
x)+\nabla_\mu\chi(\tau,\vec x)
\end{equation}
\begin{equation}
\psi'(\tau,\vec x)={\rm ~e}^{ie\chi(\tau,\vec x)}\psi(\tau,\vec x) ~~,~
\bar\psi'(\tau,\vec x)=\bar\psi(\tau,\vec x) {\rm ~e}^{-ie\chi(\tau,\vec x)}
\end{equation}
when the gauge transformation function $\chi(\tau,\vec x)$ has
periodic derivatives,
\begin{equation}
\nabla_\mu\chi(\tau=1/T,\vec x)=\nabla_\mu\chi(\tau=0,\vec x)
\end{equation}
and it is periodic up to an integer multiple of $2\pi /e$,
\begin{equation}
\chi(\tau=1/T,\vec x)=\chi(\tau=0,\vec x)+2\pi
n/e ~~,~n\in{\rm Z}
\end{equation}
The group of all gauge transformations modulo those which are strictly
periodic is $Z$, the additive group of the integers.  This is a global
symmetry.  The Polyakov loop operator transforms non-trivially under
the coset when its charge is not an integer multiple of the electron
charge,
\begin{equation}
P_{\tilde e}'(\vec
x)=P_{\tilde e}(\vec x)\cdot {\rm ~e}^{2\pi i n\tilde e/e}
\end{equation}
It can therefore be used as an operator to explore the realization of
$Z$ in the statistical model specified by the path integral
(\ref{pi}).  If the symmetry is unbroken, the loop operator averages
to zero and the system is in the confining phase.  If it is
spontaneously broken, the loop operator can have a non-zero
expectation value.  The system is then in a non-confining phase.  We
shall discuss two such phases.  One is the Debye plasma phase which is
characterized by the exponential decay in the asymptotic behaviour of
the electric two-particle potential, related to the Debye screening
length of the plasma. It is also related to the mass of elementary
excitations in a lower dimensional theory ~\cite{eric}.  The other is
what is termed a ``screening phase''.  This phase also has an electic
screening length, as well as magnetic screening.  The main difference
between these two phases is in the temperature dependence of the
screening length.  In the screening phase, the screening length
persists at zero temperature, whereas the Debye mass vanishes at zero
temperature.

The $Z$ symmetry has a physical interpretation in terms of the charge
of physical states ~\cite{gsst}.  In path integral quantization, as we
shall discuss in Section III, the temporal component of the gauge
field $A_0$ arises as a Lagrange multiplier to enforce gauge
invariance.  The projection operator which guarantees gauge invariance
in the construction of the path integral is obtained by exponentiating
the generator of infinitesimal gauge transformations and integrating
over all gauge transformations,
\begin{equation}
{\cal P}=\frac{1}{\rm const.}\int [dA_0(\vec x)]{\rm ~e}^{
\frac{i}{T} \int d\vec x~\left(
\vec\nabla A_0\cdot\vec E-A_0 e:\psi^{\dagger}\psi:
\right)}
\end{equation}
and the density matrix for the Gibbs ensemble is
\begin{equation}
\rho = {\cal P}\frac{ {\rm ~e}^{-H/T}}{Z[T]}
\end{equation}
The $Z$ transformation
\begin{equation}
A_0(\vec x)\rightarrow A_0(\vec x)+2\pi n T/e
\end{equation}
results in
\begin{equation}
{\cal P}\rightarrow {\cal P}~\cdot~{\rm ~e}^{2\pi i nQ/e}
\end{equation}
where
\begin{equation}
Q\equiv e\int d\vec x~:\psi^{\dagger}(\vec x)\psi(\vec x):
\end{equation}
is the (normal ordered) electric charge operator.  If the $Z$ symmetry
is not broken, all physical states with non-zero weight in the
thermal ensemble have quantized charges,
\begin{equation}
{\rm ~e}^{2\pi i nQ/e}~\cdot~{\cal P}\frac{ {\rm e}^{-H/T} }{Z[T]} =
{\cal P}\frac{ {\rm ~e}^{-H/T} }{Z[T]}
\end{equation}
If the $Z$ symmetry is broken, there are states contributing to the
thermal ensemble which have non-quantized charges.  Intuitively, these
can occur in a deconfined theory since the long-ranged electric fields
accompanying arbitrarily diffuse charge distributions would have
finite energy.  Such states would not be allowed in the confining
phase.  Note that, in this operator picture, the $Z$ symmetry is not a
symmetry transform of the density matrix in the usual sense.  In fact
the existence of this symmetry is related only to the question of
whether $\exp(2\pi i ~ Q/e)$ is the unit operator when operating on
the states that have non-zero weight in the density matrix.

At $T=0$, and for the physical value of the electromagnetic coupling
constant, 3+1-dimensional electrodynamics does not exhibit a confining
phase.  It is in the deconfined Coulomb phase at zero temperature and
forms a Debye plasma at finite temperature and density.  There is a
conjecture that if the electric charge of QED could be increased to
some critical value, there would be a phase transition to a chiral
symmetry breaking and confining phase ~\cite{mir}.  It is reasonable
to expect that the resulting strongly coupled system would have a
confinement-deconfinement transition at some finite temperature.

QED in 1+1 dimensions with a massive electron, i.e. the massive
Schwinger model, is confining and the $Z$ symmetry is not broken.  In
the case of the massless Schwinger model, the $Z$ symmetry is
spontaneously broken ~\cite{hnz,gsst}.  This spontaneous breaking of
$Z$ is interpreted as screening, rather than deconfinement.  A
similar situation appears in 1+1-dimensional QCD with massless quarks
{}~\cite{ik,gkms}.  With massive quarks, the question of confinement and
the confinement-deconfinement transition in the large $N$ limit of
1+1-dimensional QCD both at finite temperature ~\cite{kut,kleb,kz} and
at zero temperature ~\cite{mish} has been examined by several authors.

When the electron has a mass, parity invariant quantum electrodynamics
(QED) in 2+1 dimensions is believed to be a confining theory.  The
tree level Coulomb interaction varies logarithmically with distance.
Its entire spectrum is bound states, although the bound states can
have arbitrarily large sizes.  The perturbative self-energy of the
electron has a logarithmic infrared infinity ~\cite{sen}.  Also, when
the number of electron flavors is small enough, confinement,
accompanied by chiral symmetry breaking is believed to persist in the
limit as the bare electron mass is put to zero ~\cite{anw}.  In a
previous paper by three of us it was argued that 2+1-dimensional QED
has a phase transition from the confined to a deconfined phase at a
critical temperature ~\cite{gss}.  The phase transition is of
Berezinsky-Kosterlitz-Thouless~\cite{ber,kos} (BKT) type with
non-universal, temperature dependent critical exponents.  This is the
standard phase transition of the Coulomb gas, which is a reasonable
characterization of the thermal state of QED in the limit where the
density of thermally excited particles is low.  This is the case when
the mass of the particles is greater than the temperature.  The
Polyakov loop operators have power-law correlators in the confining
phase. In the deconfined phase there is an electric, Debye mass which
makes the Coulomb interaction short-ranged.  There is a universal
quantity associated with the BKT phase transition, the bulk modulus of
spin waves in the massless phase. This predicts the power law behavior
of Polyakov loop correlators in the confined theory.

It is interesting to ask what happens in the case where the three
dimensional gauge theory is not parity invariant, but the action
contains a topological mass term for the photon.  Of course, one would
expect that the resulting photon mass ~\cite{djt2,tpm} provides an
infrared cutoff and thereby removes the long-ranged interactions that
are responsible for confinement.  However, if the topological mass is
very small, so that the confinement scale is much larger than the
photon mass, we expect that some effects of confinement persist at
short distances.

To understand what effect topological mass has on the $Z$ symmetry,
consider the Chern-Simons (CS) action in Euclidean space
\begin{equation}
S_{CS}=i\frac{\kappa}{2}\int_0^{1/T}d\tau\int d\vec
x\epsilon_{0ij} \left(2 A_0\nabla_i A_j-A_i\dot A_j\right)
\end{equation}
The gauge transformation
\begin{equation}
A_0(\tau,\vec x)\rightarrow A_0(\tau,\vec x)+{d\over d\tau}
\left( \frac{e}{\kappa}nT\tau\right)
\end{equation}
where $e$ is the basic unit of charge of all matter fields, results in
the change of the action
\begin{equation}
S_{CS}\rightarrow S_{CS}+ineg
\end{equation}
where $g$ is the magnetic charge,
\begin{equation}
g=\int d\vec x \epsilon_{0ij}\nabla_i A_j
\end{equation}
When, as it happens on an open space such as ${\cal R}^2$ which we
consider here, the magnetic charge is not quantized, the Chern-Simons
action is not invariant under the gauge transformation unless $n=0$.
The presence of a Chern-Simons term in the action then would break the $Z$
symmetry explicitly.

On the other hand, if the space is compact, or if we impose boundary
conditions on an open space, such as the plane, so that the gauge
field can be stereographically projected onto a compact space, the
magnetic charge should obey the Dirac quantization condition,
\begin{equation}
eg= 2\pi k \label{diquc}
\end{equation}
where $k$ is an integer and $e$ is the basic unit of charge of the
matter fields.  In this case, the Chern-Simons term is invariant under
gauge transformations with non-zero winding number.  The global
symmetry is $Z$, even in the absence of matter.

The Polyakov loop operator for a basic charge (remember that, because of
the presence of monopoles, the charge is quantized) transforms as
\begin{equation}
{\rm ~e}^{ie\int_0^{1/T}d\tau A_0(\tau,\vec x)}
\rightarrow
{\rm ~e}^{ie^2/\kappa}\cdot {\rm ~e}^{ie\int_0^{1/T}d\tau A_0(\tau,\vec x)}
\end{equation}
We denote the coefficient of the Chern-Simons term as
\begin{equation}
\kappa=\frac{e^2} {2\pi}\frac{p}{q} \ ,
\label{kappa}
\end{equation}
 so that the Polyakov loop operator transforms by the phase 
$ {\rm ~e}^{2\pi iq/p}$. Depending of the ranges of $p$ and $q$, 
there are three possibilities:
\begin{itemize}
\item{}If $p/q$ is an irrational number, then the symmetry is
$Z$. In finite volume this will imply that all charges are confined
and only neutral configurations are allowed.
\item{}If $p$ and $q$ are integers so that $p/q$ is rational then
the symmetry is the finite cyclic group $Z_p$.  In this case, charges
which are not integral multiples of $p$ are confined.
\item{}If $q$ is an integer and $p=1$, there is no symmetry.
\end{itemize}

The latter condition is compatible with Gauss' law which, as we shall
see in Section III, relates the total charge and magnetic flux of a
quantum state on a compact space as
\begin{equation}
\frac{e^2}{2\pi}\frac{p}{q}g+Q=0
\end{equation}
Since $g=\frac{2\pi}{e}\cdot {\rm integer}$ and $Q={\rm integer}\cdot
e$, it is necessary that $p$ and $q$ are integers.  Furthermore, the basic
state in this system has electric charge $p\cdot e$ and magnetic
charge $q\cdot\frac{2\pi}{e}$.

In fact, we could think of the effective symmetry of the Euclidean
path integral as just enforcing the global constraints on the charges
which are contained in Gauss' law.  We shall examine the question of
whether this symmetry survives in the infinite volume limit
in Section III.

In order to analyze symmetry breaking, the properties of correlators
and other dynamical questions in parity invariant 2+1-dimensional QED,
we shall compute the effective action for the Polyakov loop operator
in Section II.  This method was advocated in the seminal work
Svetitsky and Yaffe in the context of lattice gauge
theories~\cite{sve}. One first fixes the static temporal gauge,
\begin{equation}
\frac{d}{d\tau}A_0(\tau,\vec x)=0
\end{equation}
and integrates all the degrees of freedom of the gauge theory
except for $A_0$ which is associated with the Polyakov loop.  This
generates an effective theory for the order parameter which, for an
initial $d+1 $-dimensional gauge theory is a $d$-dimensional scalar
field theory with variable $A_0(\vec x)$.  This theory exhibits the
global $Z$ symmetry explicitly.  Since, at finite temperature,
fermions always have a mass gap, integrating out all the other degrees
of freedom in the original theory produces only short-ranged
interactions in the effective theory.  Consequently, the critical
behavior of $d+1$-dimensional finite temperature QED is that of the
$d$-dimensional effective local field theory.  By studying the
effective action, we are able to characterize the type of the phase
transition and discuss the associated critical behavior.

In Section III we shall consider 2+1-dimensional pure QED with a
Chern-Simons term in the presence of an array of external charges.  We
consider the case where the space is compact and discuss some
boundary problems related to the presence of the Chern-Simons term on
a compact surface. We discuss canonical quantization and the
construction of the functional integral representation of the
partition function $Z[T]$.  We then find the exact effective action
for $A_0$ by integrating all the spatial components of the gauge
field.  We use this effective action to study the realization of $Z$
symmetry in the infinite volume limit.

In Section IV, we use a variational method to examine the
behavior of both parity invariant and topologically massive
QED in infinite volume.  We confirm the existence of the BKT
phase transition in the parity invariant theory. We also find
indications of a similar transition in topologically massive QED.
We speculate that the latter transition, although not strictly
speaking a phase transition, separates two regions of the system with
distinct physical behavior: a low temperature screening phase where 
there are charged particles bound into neutral bound states, as well
as free neutral particles, and a high temperature deconfined phase
where the bound states are absent and the charges of the particles are
Debye-screened.  Section V is devoted to a discussion of our results.

In this paper, we do not address the interesting and controversial
question of whether domain walls exist between regions with different
``orientations'' of broken $Z$ symmetry.  The existence of these in
non-Abelian gauge theories has been a subject of much discussion
{}~\cite{3.,4.,sm,kiskis}.  If they did exist in QED, they would be very
interesting and perhaps observable objects.  Detailed analysis of this
possibility is still an important problem.

It is interesting that in this paper we find a phase transition which
is accessible to perturbation theory.  This is a property of the
critical line of BKT transitions; one end of the line is in a
perturbative regime.  To our knowledge, this is the only situation
where one can study a confinement-deconfinement transition without the
aid of numerical simulations. On the other hand, confirming the
existence and properties of the transition which we discuss by
numerical simulations would be a most worthwhile project.

\section{Effective action in parity invariant QED}

We shall consider QED in 2+1 dimensions.  As it is well known, in
2+1 dimensions, the minimal, two-component Dirac fermions violate
parity ~\cite{djt2,tpm}, so that if included in the action, they can
generate a parity violating topological mass for the photon by
radiative corrections~\cite{ns,re}. In this section we shall study the
case where the electron has mass but the photon is massless.  For this
purpose, we shall use parity invariant four-component fermions which
are obtained by dimensional reduction of the 3+1-dimensional Dirac
operator.  The resulting model has two species of massive two-component
fermions where the mass terms have opposite signs.  The Euclidean
action is
\begin{equation}
S=\int d^3x\left[
\frac{1}{4}F_{\mu\nu}^2+
\bar\psi(\gamma\cdot(\nabla-ieA)+m)\psi
\right]
\label{qed}
\end{equation}
We shall find the effective action for the Polyakov loop operator
(\ref{pol}).  At finite temperature, it is possible to use a gauge
transformation to set the temporal component of the gauge field,
$A_0(\tau,\vec x)$, independent of the Euclidean time, $\tau$.  In
that gauge,
\begin{equation}
P_{\tilde e}(\vec x)={\rm ~e}^{i\tilde eA_0(\vec x)/T}
\end{equation}
and the effective field theory for this operator is the effective
action for the static field $A_0(\vec x)$.  This two-dimensional
effective field theory is obtained by integrating the other degrees of
freedom from the path integral
\begin{equation}
{\rm ~e}^{-S_{\rm eff}[A_0]}\equiv \int [d\vec Ad\psi
d\bar\psi]{\rm ~e}^{-S[A_\mu,\psi,\bar\psi]}
\end{equation}
(The functional integral defines the effective action only modulo a
temperature and volume dependent but $A_0$ independent constant.)  The
integral over the Grassman-valued Fermi fields yields the determinant
of the Dirac operator, and
\begin{equation}
S_{\rm eff}[A_0]=\int d\vec x~\frac{\vec \nabla A_0\cdot\vec\nabla
A_0}{2T} ~-~\ln\int[d\vec A]~\det[\gamma\cdot(\nabla-ieA)+m]~
{\rm ~e}^{-\int_0^{1/T}d\tau d\vec x~ F_{ij}^2/4}
\label{effac}
\end{equation}
The first term in (\ref{effac}) is the tree level action and the
second term contains all quantum corrections.  The remaining
functional integral requires additional gauge fixing.  The $Z$
symmetry is a periodicity of the effective action under the field
translation
\begin{equation}
S_{\rm eff}[A_0]~\rightarrow S_{\rm eff}[A_0+ 2\pi n T/e]
\end{equation}

The effective action, $S_{\rm eff}[A_0]$, is non-local and
non-polynomial.  It can only be regarded as a local field theory when
the momenta of interest are much smaller than the mass gaps of the
fields which have been eliminated.  In that case the effective action
has a local expansion in powers of derivatives divided by masses.

At finite temperature the action (\ref{qed}) contains three parameters
with the dimension of mass, the electron mass $m$, the gauge coupling
$e^2$, and temperature $T$. The loop expansion is super-renormalizable
{}~\cite{djt2,ap}. In fact, with a gauge invariant and Euclidean Lorentz
invariant regularization, it has no divergences whatsoever.  This
means that the effective theory that we obtain does not contain an
ultraviolet cutoff, and its {\it only} dimensional parameters are
$e^2$, $m$ and $T$.

The dimensionless parameter which governs the accuracy of the loop
expansion is the smaller of $e^2/m$ and $e^2/T$.  Also, in order for
the local, derivative expansion of the effective action to be valid,
it is necessary that either the electron mass or the temperature be
larger than the momentum scales of interest.  Then, the larger of $m$
or $T$ acts as the ultraviolet cutoff of the effective field theory.

At the tree-level, where the effective action is approximated by the
first term on the right-hand-side of (\ref{effac}) only, we can easily
compute the correlator of Polyakov loop operators,
\begin{equation}
\left\langle {\rm ~e}^{i\tilde eA_0(\vec x)}~ {\rm ~e}^{-i\tilde
eA_0(0)}\right\rangle_{\rm tree}~=~ {\rm const.}\prod_{i<j}\vert \vec
x\vert^{-\tilde e^2/4\pi T}
\label{kt}
\end{equation}
It has a scale invariant form with a temperature dependent exponent
reminiscent of the correlators in the Gaussian spin-wave theory
{}~\cite{gsw,gold}.  This is a result of the marginally confining nature of
the logarithmic Coulomb interaction.  The behavior is between that of
a confining theory where the correlator would exhibit the clustering
property and decay exponentially at large distances and the
deconfined theory where it would approach a constant.  It is
interesting to ask how quantum fluctuations would modify this result.
We shall argue in the following that, at low temperatures the behavior
(\ref{kt}) is qualitatively, though not quantitatively correct.  At
high temperatures, the correlator approaches a constant at large
distances and the $Z$ symmetry is broken.

We shall compute the effective action for $A_0(\vec x)$ in the one-loop
approximation and in an expansion in powers of derivatives of
$A_0(\vec x)$. The leading terms are
\begin{equation}
S_{\rm eff}[A_0]=\int d^2 x \left( Z(m,eA_0/{{T}})
\frac{1}{2T}\vec\nabla A_0\cdot\vec\nabla A_0 -V(m,eA_0/{T})\right)\ \ .
\label{effa}
\end{equation}
Here $V$ is the effective potential for $A_0$.  $Z-1$ arises from the
expansion of the temporal components of the vacuum polarization
function to linear order in $-\vec\nabla^2$.
\begin{equation}
\Pi_{00}(\omega=0,\vec k^2)=\Pi_{00}(0,0)+\vec k^2(Z-1)+\ldots
\end{equation}
To  one-loop order, the effective potential, $V(m,eA_0/T)$, is obtained
from the fermion determinant in the constant background $A_0$,
\begin{equation}
V(m,eA_0/T)=\frac{1}{({\rm Vol.})}
\log \det((-i\partial_0-e A_0)^2-\nabla^2+m^2)
\end{equation}
The fermions have anti-periodic boundary conditions in the compact
time.  To regularize the determinant we consider the
ratio{}~\cite{bal}
\begin{equation}
\Delta(m,eA_0/T)=\frac{\det((-i\partial_0-e
A_0)^2-\nabla^2+m^2)}{\det(-\partial^2_0-\nabla^2+m^2)}
\end{equation}
The antiperiodic boundary conditions lead to the expression
\begin{equation}
\Delta(m,eA_0/T)=\prod_{n,\vec k}\frac{((2n+1)\pi T-e
A_0)^2+k^2+m^2}{((2n+1)\pi T)^2+k^2+m^2)}
\end{equation}
The product over integers can be evaluated explicitly as ~\cite{grad}
\begin{equation}
\Delta(m,eA_0/T)=\prod_{\vec k}\left[1-\frac{\sin^2(e
A_0/2T)}{\cosh^2(\lambda_k/2T)}\right]\equiv\prod_{\vec k}\Delta_{\vec
k}(m,e A_0/T)\ \ , \label{det}
\end{equation}
where $\lambda_k^2=\vec k^2+m^2$ are the eigenvalues of the operator
$-\nabla^2+m^2$. In the infinite volume limit the $\prod_{\vec k}$ in
$\Delta(m,eA_0/T)$ gives rise to an integral on $\vec k$ in
$\log\Delta$.  The result (\ref{det}) holds in any dimensions, but the
integral on $\vec k$ can be performed analytically only in two
dimensions.  In one and three dimensions this integral can only be
done for $m=0$ in which case it gives simple polynomial expressions.
In the limit $m=0$, the effective potentials for $A_0$ have been
discussed in ~\cite{sm}.  In two dimensions we obtain
\begin{eqnarray}
V(m,eA_0/T)=\int_{-\infty}^{+\infty}
\frac{d^2\vec{k}}{(2\pi)^2}\log
\Delta_{\vec k}(m,eA_0/T)=~~~~~~~~~~~~~~~\nonumber \\
=-\frac{T^2}{\pi}\left[\frac{m}{T}Li_2({\rm ~e}^{-m/T},eA_0/T+\pi)+
Li_3({\rm ~e}^{-m/T},e A_0/T+\pi)\right]
\label{effp}
\end{eqnarray}
where
\begin{equation}
Li_2(r,\theta)=-\int_0^r dx \ln(1-2x\cos\theta+x^2)/2x
\end{equation}
and
\begin{equation}
Li_3(r,\theta)=\int_0^r dx Li_2(x,\theta)/x
\end{equation}
are the real parts of the dilogarithm and trilogarithm according to
the convention of Ref.~\cite{lew}.  As one can see from the definition
of $Li_2(r,\theta)$ and $Li_3(r,\theta)$, the effective potential is
explicitly invariant, as expected, under the $Z$ symmetry, $e A_0/T\to
e A_0/T+2\pi n$.

Computing the temporal components of the vacuum polarization function,
$\Pi_{00}(p,A_0)$,
in an external constant $A_0$ field and keeping
only the term which contributes the leading order in derivatives to
the effective action, one obtains
\begin{equation}
Z(m,e A_0/{T})=1+\frac{e^2}{12\pi m}\left(1-m\frac{\partial}{\partial
m} \right)\frac{\sinh m/T}{\cosh m/T+
\cos eA_0/{T}}
\label{effb}
\end{equation}
We will find it convenient to use the harmonic expansion of the
effective potential (\ref{effp}),
\begin{equation}
V(m,e A_0/{T})=-\frac{T^2}{\pi}\sum_{n=1}^{\infty}(-1)^n {\rm ~e}^{-n m/T}
\left(1+\frac{n m}{T}\right)\frac{\cos(n e A_0/{T})}{n^3}\ \ .
\end{equation}

There are two limits in which we can obtain more analytic information
about the effective potential.  In the regime $m>>T$, $T/m$ and
$e^2/m$ are small and $e^2/T$ is unrestricted. The higher harmonics in
the effective potential are exponentially small perturbations to the
leading term,
\begin{equation}
V(m,e A_0/T)=\frac{T m}{\pi}{\rm ~e}^{-m/T}\cos(e A_0/{T})\ \ ,
\label{sing}
\end{equation}
which is the sine-Gordon potential.  The effective field theory in
this limit is thus the sine-Gordon model,
\begin{equation}
S_{\rm eff}^{(m>>T)}[A_0]=\int d\vec x\left\{ \frac{1}{2T}
\vec\nabla A_0\cdot\vec\nabla A_0-\frac{Tm}{\pi}{\rm ~e}^{-m/T}
\cos( eA_0/T)\right\} \label{sineG}
\end{equation}
This model has a phase transition corresponding to the BKT
{}~\cite{ber,kos} transition in a two-dimensional classical Coulomb gas.
The critical behavior associated with this transition has been studied
extensively ~\cite{co,chle,sam,pasha,jose,am}.  Wiegmann ~\cite{pasha}
and Amit et al.~\cite{am} showed that in the sine-Gordon model any
perturbation of the type $\cos(n\beta\phi)$ to the sine-Gordon
potential $\alpha\cos(\beta\phi)/\beta^2$ is irrelevant to the
critical behavior of the model.  They showed that the scale dimension
of $\cos(n\beta\phi)$ is $2n^2$ at the critical point.  Consequently
for $n>1$ these operators have scale dimension greater than two and
they are indeed irrelevant.  Thus we conclude that the critical
behaviour in the regime $m>>T,e^2$ is identical to that of the two
dimensional sine-Gordon potential of eq.(\ref{sing}).

The BKT transition occurs at a line of critical points.  In the
sine-Gordon model with potential $$\alpha\cos(\beta\phi)/\beta^2$$
the critical line begins at the point
\begin{equation}
(\alpha,\beta^2)=(0,8\pi)
\end{equation}
This critical point was originally found by Coleman in his discussion
of bosonization and the correspondence of sine-Gordon theory with the
massive Thirring model ~\cite{co}.  The line of critical points in the
sine-Gordon theory corresponds to a line of critical points for the
confinement-deconfinement transition in QED which can be drawn for
example in the $\left(\frac{e^2}{m}{\rm ~e}^{-m/T},e^2/T\right)$ plane. 
Comparing eq. (\ref{sing}) with the sine-Gordon potential we see
that the QED critical line starts at
\begin{equation}
\left( \frac{e^2}{m}{\rm ~e}^{-m/T},{\rm ~e}^2/T\right)=(0,8\pi)
\end{equation}
The critical temperature corresponding to this point is
\begin{equation}
T_{\rm crit.}^{(m>>T)}=e^2/8\pi
\end{equation}
The renormalization group was used to study this phase transition,
originally by Kosterlitz ~\cite{kos} and Wiegmann ~\cite{pasha} and
later improved to higher order by Amit and collaborators ~\cite{am}.
The flow diagram is depicted in Fig. 1.  There are three regions: The
high temperature, deconfined region III, where the model is
asymptotically free.  This is the case which was analyzed by Coleman
{}~\cite{co}.  The low temperature region I is confining and has a line
of infrared stable fixed points at $m=\infty$, corresponding to $c=1$
conformal field theory.  In Region II, the model is deconfined and is
neither asymptotically nor infrared free.  The separatrix between
regions I and II is the line of BKT phase transitions.

To compute the leading correction to $T_{\rm crit.}$ due to a finite
(but still large) value of the fermion mass, one has first to
renormalize the field $A_0$, according to
\begin{equation}
A_0^{\rm ren}\equiv A_0 Z(m,0)^{1/2}
\end{equation}
This renormalization changes the sine-Gordon parameter $\beta$ in the
argument of the cosine. The one-loop calculation of the effective action
results in the correction
\begin{equation}
T^{(m>>T)}_{\rm crit.}=\frac{e^2}{{8\pi}(1+{e^2}/{12\pi
m}+\ldots)}~~~.
\end{equation}

As we have shown, the BKT phase transition in QED is a
confinement-deconfinement transition.  In the spin wave plus Coulomb
gas description of the XY-model ~\cite{gold}, the BKT phase transition
corresponds to a binding-unbinding of vortices. In QED it has the
obvious analog of a binding-unbinding transition for charged
particle-antiparticle pairs.  In the deconfined phase, $A_0$
fluctuates near one of the minima of the effective potential,
\begin{equation}
\left\langle A_0\right\rangle  ={2\pi n T}/{e}
\end{equation}
In a semiclassical analysis, this expectation value contributes an
imaginary chemical potential for the electron.  However, this chemical
potential can be absorbed by shifting the Matsubara frequency by $n$
units.  Thus, in a semiclassical analysis, the thermodynamics in the
deconfined phase does not suffer from the difficulties of negative
entropy and imaginary thermodynamic potential that affect the
meta-stable $Z_N$ phases of QCD{}~\cite{cds,bksw}.

The other limit where the one-loop result is simple is the high
temperature limit, where $T>>m,e^2$.  In that limit we must be careful
to study the degree of freedom which is periodic \footnote{ We
disagree with the discussion on this point in ref. ~\cite{sm}.}.  For
this, we define the field
\begin{equation}
a(\vec x)\equiv e A_0(\vec x)/T
\end{equation}
so that the effective action for the field $a(\vec x)$ has the periodicity
\begin{equation}
a(\vec x)\rightarrow a(\vec x)+2\pi n
\end{equation}
The effective action in the high $T$ limit is
\begin{equation}
S_{\rm eff}^{(T>>m)}[a]=\int d\vec x \left\{\frac{T}{2e^2}\vec
\nabla a\cdot\vec\nabla a+\frac{T^2}{\pi}Li_3(1,a+\pi)\right\}
\end{equation}
Large $T$ is the semi-classical limit for this theory and $a$ must
fluctuate near a minimum of the effective potential.  In this case, as
expected, the $Z$ symmetry is spontaneously broken, corresponding to
deconfinement.

\section{Maxwell-Chern-Simons theory on the sphere}

It is interesting to ask what happens in three-dimensional QED when it is
not parity invariant.  In this case, the gauge field can have a
topological mass term ~\cite{djt2} and, naively, one would expect that
confinement is not an issue, it is simply absent.  In this Section, we
shall consider the properties of finite temperature
Maxwell-Chern-Simons theory when the space is the 2-sphere.  We shall
find that there is an analog of the $Z$ symmetry, which exists and has
interesting properties even in the absence of matter fields.  The
symmetry enforces a kind of topological confinement which arises from
Dirac's quantization condition for the magnetic field of the monopole.
For completeness, we shall also give a careful treatment of the finite
temperature path integral in this case.  The Minkowski space action
for Maxwell-Chern-Simons theory with Wilson-loop sources is
\begin{equation}
S=\int d^3x\left\{ -\frac{1}{4}F_{\mu\nu}F^{\mu\nu}+
\frac{\kappa}{2}\epsilon^{\mu\nu\lambda} A_\mu \partial_\nu A_\lambda
\right\}+\sum_ie_i \int_{\Gamma_i} dx^\mu A_\mu
\label{action}
\end{equation}
The spacetime is a product of time $R^1$ and $S^2$.  A similar
construction can be carried out where the space is a product of $R^1$
and any Riemann surface with some additional complications
{}~\cite{berg}.

Since the space is compact and we shall consider the situation where
the total charge of the external charge distribution is not zero, the
Gauss' law constraint will force us to consider the case where there
is a non-zero magnetic charge
\begin{equation}
\int_{S^2} dA=g \neq 0
\end{equation}
In this case, the spatial components of the gauge field are not
globally defined functions on the sphere but are rather components of
a connection on the monopole line bundle.  It is well known that, in
this situation, extra topological terms are needed to make the
Chern-Simons term well-defined ~\cite{alvarez,berg}.  This arises from
the fact that the density which one integrates to get the Chern-Simons
term is not gauge invariant, but transforms by an exact form under
gauge transformations.  This makes the integral of the Chern-Simons
three-form sensitive to the coordinatization of the monopole line bundle.

This sensitivity can be seen by the following argument.  We decompose
the gauge field into its spatial and temporal part,
\begin{equation}
A_\mu dx^\mu=A_0 dt+A_i dx^i\equiv A_0 dt+A
\end{equation}
We construct the monopole line bundle by considering a set of
coordinate patches, $\{P_k\}$, which cover the 2-sphere,
\begin{equation}
\cup_k P_k =S^2
\end{equation}
and denoting the gauge field in the $k$'th patch as $A^k$.  The gauge
field in different patches are related by gauge transformations,
\begin{equation}
A^k-A^p=d\chi^{kp}
\end{equation}
where $\chi^{kp}$ is a function which is defined on the intersection
region $P^k\cup P^p$.  The integral over $S^2$ is defined using a
partition of overlapping patches, so that each point of $S^2$ is
integrated only once.  Gauge invariant quantities such as $dA$ or
$dA\*dA$ are not sensitive to details of the choice of coordinate
patches.  Likewise, the integral of gauge invariant quantities 
does not depend on the positions of the patches.  However, the
integral of a non-gauge invariant density, such as the (naive)
Chern-Simons term
\begin{equation}
\int
dt\sum_{k}\int_{P_k}\left( A_0 dA^k+ A^kdA_0-A^k\frac{d}{dt}A^k\right)
\end{equation}
depends on the position of the patches. For example, the contribution to
Gauss' law density arising from the Chern-Simons term is obtained by
taking a variational derivative of the Chern-Simons term by $A_0$.  To
do this, it is necessary to translate $A_0 \rightarrow A_0+\delta
A_0$, to isolate $\delta A_0$ by integrating by parts and then to
identify the functional derivative as the coefficient of $\delta A_0$
under the integral.  With this procedure one obtains the charge
density
\begin{equation}
J_0= 2dA+\sum_{kp}\int_{P_k\cap P_p}\delta(x-P_k\cap P_p)
d\chi^{kp}
\end{equation}
This has the unappealing feature that some charge lives on the
arbitrarily chosen transition regions (in fact, on the arbitrarily
chosen boundaries between patches which are inside the transition
regions).  The surface terms which must be added to the Chern-Simons
term cure this difficulty.  In the present case, they are integrated
on the intersection regions of coordinate patches and cancel the terms
obtained when the second term in the above naive Chern-Simons term is
integrated by parts to obtain the expression
\begin{equation}
\int dt\sum_k
\int_{P_k}\left( 2A_0dA^k-A^k\dot A^k\right)
\end{equation}
This version of the Chern-Simons term will be sufficient for our
purposes in the following.

\subsection{Canonical quantization}

We shall examine the electrostatics of an array of classical charges.
In this case, the source term in the action has the form
\begin{equation}
\sum_i e_i \int dt A_0(x_i,t)
\end{equation}
In particular, we are interested in the free energy of this system as
a function of particle positions.  We shall not require global
neutrality of the charge distribution. However, the consistency of the
monopole bundle will force us to use charges which are integer
multiples of a basic charge, $e$, compatible with the Dirac
quantization condition.

If the total charge is non-zero, the gauge constraint, which is
obtained by taking a functional derivative of the action
(\ref{action}) by $A_0$ is
\begin{equation}
\nabla\cdot E +\kappa B +\sum_i e_i \delta(x-x_i)\sim0
\end{equation}
The electric field is gauge invariant and must be a globally defined
vector field on $S^2$.  Therefore, the integral of the divergence of
the electric field over the space $S^2$ must vanish. The integral of
Gauss' law then implies that, when the total electric charge is not
zero, there is also a non-zero magnetic flux,
\begin{equation}
\kappa g+\sum_i e_i=0
\end{equation}

It is convenient to separate the effect of the magnetic background
field by decomposing the gauge field into a classical time-independent
part containing the monopole field and a time dependent part with no
overall magnetic flux and which is allowed to have quantum
fluctuations,
\begin{equation}
A_i(x,t)\rightarrow A_{M,i}(x)+ A_i(x,t)\ \ .
\label{mon}
\end{equation}
In eq.(\ref{mon})
\begin{equation}
\int_{S^2} \nabla\times A_M=g
\end{equation}
and $A_M$ is defined in such a way that the classical magnetic field
is constant,
\begin{equation}
B_M=\nabla\times A_M=g/4\pi R^2
\end{equation}
(with $R$ the radius of $S^2$) so that $$\int B=0$$ and $$\int
B_MB=0~~~.$$

Substituting (\ref{mon}) into the action yields
\begin{eqnarray}
S=\int dt\int_{S^2}\left\{ \frac{1}{2}\left(\dot
A_i-\nabla_iA_0\right)^2 -\frac{1}{2}(B_M^2+B^2) +\kappa A_0B_M+\kappa
A_0B
\right.  \nonumber\\  \left.
-\frac{\kappa}{2}A\times \dot A
-\frac{\kappa}{2}\frac{d}{dt}\left( A_M\times A\right)\right\}+
\sum_i\int dt e_i A_0(t,\vec x_i)
\end{eqnarray}
where, now all variables are single-valued, globally defined functions
on $S^2$ and the monopole gauge field $B_M$ is a classical variable.
The only remaining multi-valued term $\int {d\over dt}(A_M\times A)$
is a total time derivative term and therefore is not important for the
canonical quantization which we shall do in the following.

Canonical quantization proceeds by identifying the  canonical momenta
\begin{eqnarray}
\Pi_0&\sim0&
\label{const}\\
\Pi_i&=&\dot A_i-\nabla_i A_0+\frac{\kappa}{2}
\epsilon_{ij}A_j
\end{eqnarray}
The first relation is a primary constraint.  The Hamiltonian is
\begin{equation}
H=\int_{S^2}\left\{
\frac{1}{2}\left(\Pi_i-\frac{\kappa}{2}\epsilon_{ij}A_j
\right)^2+\frac{1}{2}B^2+\frac{1}{2}B_M^2\right\}
\label{ham}
\end{equation}
and the canonical commutation relations are given by
\begin{eqnarray}
\left[ A_i(\vec x),\Pi_j(\vec y)\right]&=&i\delta_{ij}\delta(\vec x-\vec y)
\label{com}
\\
\left[ A_0(\vec x),\Pi_0(\vec y)\right]&=&i\delta(\vec x-\vec y)
\end{eqnarray}
The gauge constraint arises as a secondary constraint from requiring
that the primary constraint $\Pi_0\sim0$ is time-independent
\begin{equation}
G(\vec x)\equiv
\vec \nabla\cdot\vec \Pi(\vec x)+\frac{\kappa}{2}B(\vec x)+
\kappa B_M +\sum_i
e_i\delta(\vec x-\vec x_i)\sim0
\label{gauss}
\end{equation}

The operator $G(\vec x)$ generates time-independent gauge
transformations and commutes with the Hamiltonian which is invariant
under the gauge transformation
\begin{eqnarray}
\vec A(\vec x) \rightarrow \vec A(\vec  x)+\left\{
\vec A(\vec x),\int d\vec y ~
\chi(\vec y)G(\vec y)\right\}&=&A(\vec x) -\vec \nabla\chi(\vec x)
\nonumber \\
\vec \Pi(\vec x)\rightarrow\vec \Pi(\vec x)+
\left\{\vec \Pi(\vec x),\int d\vec y~
\chi(\vec y)G(\vec y)\right\}&=&\vec  \Pi(\vec x)-\frac{\kappa}{2}
\vec \nabla^*\chi(\vec x)
\end{eqnarray}
where $\vec \nabla^*_i\equiv \epsilon_{ij}\nabla_j$.

The dynamical system with Hamiltonian (\ref{ham}) and canonical
commutator (\ref{com}) is internally consistent and can be quantized
as it is (with the subtlety that the ground state is not
normalizable).  The primary constraint, (\ref{const}) is solved by
imposing the auxiliary gauge fixing condition
\begin{equation}
A_0(\vec x)\sim0
\end{equation}
and thereby eliminating both $A_0$ and $\Pi_0$.  The Gauss' law
constraint (\ref{gauss}) must then be imposed as a physical state
condition.  Since the operator $G(\vec x)$ commutes with the
Hamiltonian, it can be diagonalized simultaneously with the
Hamiltonian.  Then the simultaneous eigenstates of the Hamiltonian and
$G(\vec x)$ which are in the kernel of $G(\vec x)$ are chosen as the
physical states,
\begin{equation}
G(x)\vert~{\rm physical~state}>=0\ \ .
\end{equation}
We can form a projection operator onto physical states by considering
the set of all gauge transforms.  Formally,
\begin{equation}
{\cal P}=\frac{1}{\rm vol.G}\int
[d\chi(x)]~\exp\left(i\int_{S^2}\chi(x)G(x)\right)\ \ .
\end{equation}
This operator satisfies the property $$ {\cal P}^2~=~ {\cal P} $$
(This is a formal statement due to the infinite volume of the gauge
group.  This is the same difficulty which appears in the normalization
of the states.)

The form of Gauss' law indicates that the gauge symmetry is realized
in a projective representation.  For example, if we represent the
canonical commutation relation in the functional Schr\"odinger
picture, where states are wave-functionals $\Psi[\vec A]$ of the
classical field $\vec A$, and the canonical momentum is a functional
derivative operator,
\begin{equation}
\Pi_i(\vec x)=\frac{1}{i}\frac{\delta}{\delta
A_i(\vec x)} \ \ .
\end{equation}
Then a physical state which obeys the gauge condition
\begin{equation}
G(\vec x)\Psi_{\rm phys}[\vec A]=0
\end{equation}
gauge transforms as
\begin{equation}
\Psi_{\rm phys}[ \vec A-\vec \nabla\chi]= {\rm ~e}^{i{\kappa}\int
\chi (B_M+B/2)+i\sum_ie_i\chi(\vec x_i)}\Psi_{\rm phys}[\vec A]
\label{proj}
\end{equation}
In the next subsection, we shall find the functional integral
representation of the thermodynamic partition function.

\subsection{Functional integral representation of the partition function}

In this subsection, we shall discuss the derivation of the functional
integral representation of the thermodynamic partition function.  It
is obtained by taking the trace over physical states of the Gibbs
distribution operator
\begin{equation}
\rho={\rm ~e}^{-H/T}
\end{equation}
where $H$ is the Hamiltonian and $T$ is the temperature.  (We work in
a system of units where the Boltzmann constant, the speed of light and
Planck's constant are equal to one.)  We shall consider the
unconstrained space of states which represent the canonical
commutation relation (\ref{com}) and insert into the trace a
projection operator which projects onto the physical states:
\begin{equation}
Z[T]=\sum_s <s\vert {\rm ~e}^{-H/T}{\cal P}\vert s>
\end{equation}
Explicitly, this can be written as
\begin{equation}
Z[T]=\frac{1}{\rm vol~G} \int [d\chi(\vec x)] \int [da_i(\vec x)] < a\vert
{\rm ~e}^{-H/T} {\rm ~e}^{i\kappa\int
\chi(B_M+B/2)+i\sum e_i\chi(\vec x_i)} \vert a+d\chi>
\label{part1}
\end{equation}
where we have taken the trace using the eigenvectors of the
``position'' operator $\vec A$,
\begin{equation}
A_i(\vec x)\vert a>=a_i(\vec x)\vert a>
\end{equation}
The integration over $\chi$ in (\ref{part1}) projects the trace onto
gauge invariant states.  Since the Hamiltonian is gauge invariant, it
is sufficient to insert the projection operator once.  The field
$\chi(\vec x)$ is proportional to the temporal component of the gauge
field, $A_0(\vec x)$, in the  time-independent gauge
\begin{equation}
\chi(\vec x)\equiv A_0(\vec x)/T \ \ .
\end{equation}

We are particularly interested in deriving an effective action for the
gauge function $A_0(\vec x)$.  It is defined by
\begin{equation}
{\rm ~e}^{-S_{\rm eff}[A_0]}\equiv \sum_{B_M} \int [d\vec a] <\vec a\vert {\rm ~e}^{-H/T}
{\rm ~e}^{i\frac{\kappa}{T}\int A_0(B_M+B/2)}\vert\vec a+\vec \nabla A_0/T>
\label{part2}
\end{equation}
where we have omitted the external charges and we have summed over
magnetic monopole number.  The subsequent integration over $A_0(\vec
x)$, which is needed in order to obtain the partition function of the
system in the presence of external charges, will enforce Gauss' law.
In particular it will project onto the sector with the correct
magnetic charge.

The partition function of the system in the presence of external
charges is the correlator
\begin{equation}
Z[T,(e_i,\vec x_i)]= \frac{
\int [dA_0(\vec x)] {\rm ~e}^{-S_{\rm eff}[A_0]} ~\prod_j {\rm ~e}^{ ie_jA_0(\vec x_j)/T} }
{ \int [dA_0(\vec x)] {\rm ~e}^{-S_{\rm eff}[A_0]} }
\label{part3}
\end{equation}

The matrix element in (\ref{part1}) has the standard phase space path
integral expression ~\cite{schul}, so that
\begin{equation}
{\rm ~e}^{ -S_{\rm eff}[A_0] } \equiv \int[d\vec a][d\vec \pi]
{\rm ~e}^{ \int_0^{1/T} d \tau \int_{S^2} \left\{ i\vec\pi\cdot\dot{\vec a}-
\frac{1}{2}(\pi_i-\frac{\kappa}{2}\epsilon_{ij}a_j)^2
-\frac{1}{2}(B_M^2+b^2) \right\} + i\frac{\kappa}{T}
\int_{S^2}A_0 \left( B_M+b/2 \right) }
\label{part4}
\end{equation}
where the time interval is $\tau\in[0,1/T]$, the spatial integral in
the action is taken over $S^2$, the canonical momentum $\pi$ has open
boundary conditions and the gauge field has the boundary condition
which is periodic up to a twist by a gauge transformation,
\begin{equation}
\vec a(1/T,\vec x)=\vec a(0,\vec x)-\vec\nabla A_0(\vec x)/T
\end{equation}
We have denoted the fluctuating part of the magnetic field as $b=\vec
\nabla\times\vec a$.  Equation (\ref{part4}) gives the effective action
up to an overall temperature dependent but $A_0$ independent constant.

The canonical momentum can be integrated in order to present the
functional integral in configuration space, up to an irrelevant
temperature dependent factor,
\begin{equation}
{\rm ~e}^{-S_{\rm eff}[A_0]}=
\sum_{B_M}\int[d\vec a]{\rm ~e}^{-\int_0^{1/T}d\tau\int_{S^2}\left(
\frac{1}{2}\dot {\vec a}^2+\frac{1}{2}(B_M^2+b^2)+i\frac{\kappa}{2}
\vec a\times\dot {\vec a}\right)} \cdot
{\rm ~e}^{i\frac{\kappa}{T}\int_{S^2}A_0(B_M+b/2)}
\end{equation}

It is convenient to untwist the boundary condition using the change of
variables,
\begin{equation}
\vec a(\tau,\vec x)\rightarrow
\vec a(\tau,\vec x)-\tau\vec \nabla A_0(\vec x)
\end{equation}
with the result that
\begin{eqnarray}
{\rm ~e}^{-S_{\rm eff}[A_0]}=\sum_{B_M} {\rm ~e}^{-\frac{1}{2T}\int_{S^2}(\nabla
A_0)^2+ i\kappa\int A_0 B_M/T}\cdot~~~~~~
\nonumber\\
\cdot\int[d\vec a] {\rm ~e}^{-\int_0^{1/T}
d\tau\int_{S^2}\left( \frac{1}{2}\dot {\vec a}^2
+\frac{1}{2}B_M^2+\frac{1}{2}b^2+i\frac{\kappa}{2}\vec a\times\dot
{\vec a}-i\kappa A_0 b\right)}
\label{part5}
\end{eqnarray}
The Gaussian integral over $\vec a_i$ can now easily be done.  If we
choose $B_M$ to be a constant magnetic field on the sphere, the
integral over $\vec a(\vec x)$ in (\ref{part5}) yields
\begin{eqnarray}
{\rm ~e}^{-S_{\rm eff}[A_0]}=\sum_{B_M} \exp\left( -\frac{g^2}{2T(4\pi
R^2)}+i\frac{\kappa g}{(4\pi R^2)T}\int_{S^2}A_0-
\right. \nonumber\\ \left.~~~~~~~~~~-
\frac{1}{2T}\sum_{l,m;l\neq0} \vert A_0(l,m)\vert^2(l(l+1)/R^2+
\kappa^2)\right)
\label{eff}
\end{eqnarray}
where we have dropped an irrelevant infinite constant, recalled the
fact that $\int_{S^2}B_M=g=4\pi R^2B_M$ and we have expanded $A_0(\vec
x)$ in spherical harmonics,
\begin{equation}
A_0(\vec x)=\sum_{lm} A_0(l,m) Y_{lm}(\vec x)
\end{equation}
$l\geq 0$ are integers of the usual angular momentum spectrum, the
spectrum of the laplacian $-\nabla^2$ being $l(l+1)R^2$, and
$m=-l,-l+1,\ldots,l\ $ for each $l$.  Notice that the non-zero modes of
the gauge function $\chi(\vec x)$ are governed by a massive euclidean
free field theory.  The zero modes, on the other hand are coupled to
the monopole moments which must be summed.

If we recall the monopole quantization condition,
$g=2\pi n/e$, the summation over monopoles is
\begin{equation}
\sum_n \exp\left( -\frac{(2\pi)^2}{(4\pi R^2)2e^2T}n^2+
\frac{2\pi i\kappa}{eT}n\hat A_0\right)
\end{equation}
where $$
\hat A_0\equiv \frac{1}{4\pi R^2}\int_{S^2}A_0
$$ Using the Poisson resummation formula, the sum can be presented as
\begin{equation}
\sqrt{\frac{(4\pi R^2)e^2T}{2\pi}}\sum_n\exp\left(
-\frac{ (4\pi R^2)e^2T}{2}\left( \frac{ \kappa}{eT}\hat A_0
+n\right)^2\right)
\end{equation}
which explicitly exhibits periodicity in $\hat A_0$,
\begin{equation}
\hat A_0\rightarrow \hat A_0 + \frac{eT}{\kappa}
\label{zs}
\end{equation}
The summation over monopoles effectively represents this symmetry in
the Villain form ~\cite{vil},
\begin{equation}
{\rm ~e}^{-S_{\rm eff}[A_0]}=\sum_n
\exp\left(-\frac{1}{2T}\int_{S^2}\left\{
\vec\nabla A_0\cdot\vec\nabla A_0
+\kappa^2\left(A_0+\frac{eT}{\kappa}n\right)^2 \right\} \right)
\label{tmea}
\end{equation}
This action is identical (although one dimension higher) to the action
that was found for the Schwinger model in ref. ~\cite{gsst}.  In that
case, we argued that the $Z$ symmetry is always broken in the infinite
volume limit.  This symmetry breaking is interpreted as screening,
similar to that which occurs in a Higgs phase, rather than
deconfinement.  We conjecture that a topologically massive gauge
theory screens, rather than deconfines.  We shall derive support for
this conjecture from the variational calculation in Section IV where
we find indications of a rapid change of the behavior of the system
between what we here interpret as a screening phase and what we would
properly call a deconfined phase.

In the next subsection we shall examine the consequences of this
symmetry of the effective action (\ref{tmea}) and we will show that it
is also spontaneously broken in the infinite volume limit.

\subsection{Spontaneous breaking of $Z$ symmetry}

The Polyakov loop operator transforms under (\ref{zs}) as
\begin{equation}
{\rm ~e}^{ineA_0(\vec x)/T}~\rightarrow ~{\rm ~e}^{in e^2/\kappa}~
{\rm ~e}^{ineA_0(\vec x)/T}
= {\rm ~e}^{2\pi i n\cdot p/q }~{\rm ~e}^{ineA_0(\vec x)/T}
\end{equation}
where we have defined (as in (\ref{kappa}) )
$$\kappa=\frac{e^2}{2\pi}\frac{p}{q}$$ This symmetry has implications
for the correlator of Polyakov loop operators,
\begin{equation}
Z[T,(e_i,x_i)]= \left\langle  \prod_j {\rm ~e}^{ie_iA_0(\vec x_j)/T}\right\rangle 
\label{fff}
\end{equation}
which depend on the ratio $p/q$.

\begin{itemize}
\item{}p/q is irrational.  Then the correlator (\ref{fff}) vanishes unless
$\sum_i e_i=0$.
\item{}p/q is rational.  Then the correlator (\ref{fff})
vanishes unless $\sum_je_j= e\cdot {\rm integer}\cdot q $.  Since the
consistency with the monopole bundle requires that the charges are
quantized in units of $e$, so $\sum_j e_j={\rm integer}\cdot e$, the
$Z$ symmetry here is actually a subgroup of $Z$, $Z_q$, the additive
group of the integers modulo $q$.
\item{}If $p$ is an integer and $q=1$ the correlator (\ref{fff})
is unrestricted.
\end{itemize}

It is interesting to observe that only charges which are integer
multiples of $q e$ are allowed on the sphere.  This is no surprise as it
is a direct consequence of Gauss' law.  The integral of Gauss' law
together with the Dirac quantization condition yields the constraint 
$\mbox{~charge}= q e \cdot{\rm ~integer~}$.  Thus, only charges which are
integer multiples of $q$ are consistent with the Gauss' law constraint
on the sphere.  The $Z_q$ symmetry enforces this ``topological
constraint''.

The $Z_q$-symmetry is invariably broken in the infinite volume limit.
To see this, we consider the two-point correlator of loop operators
\begin{equation}
\left\langle {\rm ~e}^{ieA_0(\vec x)/T}~{\rm ~e}^{-ieA_0(\vec 0)/T}\right\rangle =
{\rm const.}\cdot \exp\left( - \frac{e^2}{2T}(\vec
x\vert\frac{1}{-\vec\nabla^2+\kappa^2} \vert \vec 0)\right)
\end{equation}
If we take the limit as the volume goes to infinity, followed by the
limit as the separation of the points in the correlator goes to
infinity, the two-point function approaches a non-zero constant,
\begin{equation}
\lim_{\vert\vec x\vert\rightarrow\infty} \lim_{R^2\rightarrow\infty}
\left\langle {\rm ~e}^{ieA_0(\vec x)/T}~{\rm ~e}^{-ieA_0(\vec 0)/T}\right\rangle =
{\rm const.}
\end{equation}
This implies that the $Z_q$ symmetry is spontaneously broken in the
infinite volume limit.

This leaves us with the correct conclusion that the topologically
massive gauge theory is not a confining theory.  In fact its
electrostatic interactions are short-ranged and Yukawa-like. Their
large distance fall-off is governed by the inverse topological mass.

Thus, for all practical purposes, the topological mass of the photon
contributes a mass term to the effective action for $A_0$.  The fact
that this mass term is periodic when the volume is finite is
irrelevant in the infinite volume limit.  If we couple topologically
massive QED to matter fields, we would expect that, in the limit where
the matter field masses are large, the effective field theory is the
massive sine-Gordon theory,
\begin{equation}
S_{\rm eff}^{m>>T}[A_0]=\int d^2x\left\{ \frac{1}{2T}\vec\nabla A_0\cdot
\vec\nabla A_0+\frac{\kappa^2}{2T}A_0^2-\frac{mT}{\pi}{\rm ~e}^{-m/T}\cos(eA_0/T)
\right\}
\end{equation}
In the next Section, we shall study this model using a variational
approach.

\section{Variational approach}

In this section, we shall discuss a variational approach to the
problem of showing the existence of a phase transition in the one-loop
effective theory.  We have argued that the effective field theory
where the phase transition can be studied is the sine-Gordon theory
with an additional mass term for the boson,
\begin{equation}
S_{\rm eff}=\int d^2x\left\{ \frac{1}{2}\nabla\phi\cdot\nabla\phi
+\frac{\mu^2}{2}\phi^2-\frac{\alpha}{\beta^2}\cos\beta\phi\right\}
\label{sg}
\end{equation}
Here, the mass term for the boson is the topological photon mass,
$\mu=\kappa$.  Also, in the limit where the fermion masses of QED are
much greater than the temperature and charge squared, we have
$\beta=e/\sqrt{T}$ and $\alpha=e^2m{\rm ~e}^{-m/T}/\pi$.  The ultraviolet
cutoff is the fermion mass, $m$.

As we have discussed in the previous sections, the mass term for the
boson should really be a periodic one.  However, quantum effects
invariably break the translation symmetry for such a Bose field in the
limit where the volume is infinite.  We shall therefore ignore the
translation symmetry.

In Ref. \cite{fro}, the behavior of the model (\ref{sg}) was
investigated around the point $\beta^2=4\pi$, and a phase transition
at the corresponding critical temperature $\tilde T_c=e^2/4\pi$ was
suggested. In subsequent work on the massless model \cite{am}, it was
however found that the divergencies at $\beta^2=4\pi$ can be handled
within the renormalization scheme, and the behavior is smooth.
Accordingly, we shall concentrate on studying the model only in the
neighborhood of $\beta^2=8\pi$.

In Section II we reviewed the argument of ref.~\cite{gss} that, when
the topological mass is zero, there is a
Berezinskii-Kosterlitz-Thouless confinement-deconfinement transition
at the temperature $T_c[e^2,m]=e^2/8\pi(1+...{\cal O}(e^2/m))$,
corresponding to $\beta^2\simeq 8\pi$ for the sine-Gordon parameter.
When there is a topological mass, the system always fails to attain
its asymptotic scaling regime, and the flow diagram cannot be that of
a BKT transition.  In fact, at a scale of the order of the inverse
topological mass there is a crossover from a behavior governed by the
long ranged Coulomb interaction to one governed by a finite range
Yukawa interaction.  There is no confinement in the strict sense,
since the Coulomb interaction cuts off at a given distance.  However,
when the topological mass is small, much smaller than any of the other
dimensional parameters (particularly the confining scale which is
governed by $e^2$), for low temperature and spatial scales much less
than the inverse topological mass and of the order of the confinement
scale, the physical behavior of the system should be much like a
confining one with a confining interaction between oppositely charged
particles.  One expects that, at or near the confinement-deconfinement
temperature, a drastic change in the properties of the system takes
place, though not a true phase transition. Such a rapid change is
between a quasi-confining and a de-confined behavior with the electric
mass of the photon moving from the value of the topological mass, in
the low temperature quasi-confined phase, to the value of the much
larger Debye screening mass, in the de-confined phase.

If one imagines separating a test particle-antiparticle pair at short
distances the potential energy should increase with distance just as
it does in a strictly confining system.  If this increasing energy
gets large enough it can produce a pair of dynamical charges which
partially screen the charges of the source.  Thus, at short distances,
separating a particle-antiparticle pair should produce mesons, rather
than free charged particles.  However, at distances larger than the
inverse of the topological mass, charged particles should behave as
free particles.  We shall interpret this as a screening (as distinct
from deconfined) phase analogous to what happens in the Schwinger
model or massless two-dimensional QCD.

If we then increase the topological mass to the confining scale, the
system should go continuously to one which deconfines at all
scales. We shall qualitatively study these behaviours by means of a
variational approach in the next subsections.

\subsection{Jensen's inequality}

In statistical mechanics, a variational approach uses Jensen's
inequality.  First, we shall give a brief review of this inequality
and its derivation.  Consider a statistical system with Hamiltonian
$H_1$ which we want to study the statistical mechanics of, but are
unable to solve for the sum over states to obtain the partition
function or the correlators exactly.  We consider another test
Hamiltonian, $H_0$ which contains some parameters and with which we
can solve for the partition function and correlation functions of
observables analytically.  Then, consider the Hamiltonian which
interpolates linearly between them,
\begin{equation}
H_\lambda= \lambda H_1+(1-\lambda)H_0
\end{equation}
and the free energy
\begin{equation}
W(\lambda)=-\ln \sum_s
{\rm ~e}^{-H_\lambda(s)}
\end{equation}
It has the properties
\begin{eqnarray}
{\partial W(\lambda)\over\partial\lambda}~=~\left\langle
H_1-H_0\right\rangle_\lambda {}~~~~~~~~~~~~~~\nonumber \\ {\partial^2
W(\lambda)\over\partial\lambda^2} =-\left\langle
(H_1-H_0)^2\right\rangle_\lambda+\left\langle
H_1-H_0\right\rangle_\lambda^2~\leq~0
\end{eqnarray}
where $\left\langle ...\right\rangle_\lambda$ is the expectation value
in the ensemble with Hamiltonian $H_\lambda$.  Since the curvature of
$W(\lambda)$ as a function of $\lambda$ is always less than or equal
to zero, $W(\lambda)$ obeys the inequality
\begin{equation}
W(\lambda)\leq
W(0)+\lambda\cdot\left({\partial
W(\lambda)\over\partial\lambda}\right)_{\lambda=0}
\end{equation}
which, evaluated at $\lambda=1$ is Jensen's inequality
\begin{equation}
W(1)\leq W(0)+\left\langle H_1-H_0\right\rangle_0
\end{equation}
This establishes an upper bound on the free energy of the system of
interest by the system in the variational ansatz which can be
optimized by adjusting the parameters of the variational ansatz. The
bound is saturated only when the ensembles are identical ($H_1=H_0$).

\subsection{Variational computations}

We are faced with the problem of computing the thermodynamics of the
system with Hamiltonian function
\begin{equation}
H_1=\int d^2x\left\{
\frac{1}{2}\partial_\mu\phi\partial_\mu\phi+\frac{\mu^2}{2}\phi^2
-\frac{\alpha}{\beta^2}\cos\beta\phi\right\}
\end{equation}
We shall study this system variationally by beginning with the test
ensemble governed by the quadratic Hamiltonian
\begin{equation}
H_0=\int d^2xd^2y \frac{1}{2}\phi(\vec x)\xi((\vec x-y)^2)\phi(y)
\end{equation}
with an ultraviolet cutoff $\Lambda$, which is of order the mass of
the matter field in the original model.  The Gaussian functional
integrals appearing on the right-hand side of Jensen's inequality
in this case can easily be done with result
\begin{equation}
\frac{1}{V}W\leq\frac{1}{2}\int^\Lambda
\left\{\ln\left(\frac{\xi(p^2)}{\Lambda^2}\right)
+\frac{p^2+\mu^2}{\xi(p^2)}-1 \right\}-\frac{\alpha}{\beta^2}
{\rm ~e}^{-\frac{\beta^2}{2}\int^\Lambda \xi^{-1}}\ \ .
\label{var}
\end{equation}
Here the measure for the cutoff momentum integration is
\begin{equation}
\int^\Lambda\equiv
\int\frac{d^2p}{(2\pi)^2}\theta(\Lambda^2-p^2)
\end{equation}
and has been extracted a factor of the spatial volume to get the free
energy density.

In order to optimize the ansatz, we take the variational derivative of
the right hand side to obtain
\begin{equation}
  0=\xi(p^2)^{-2}\left(\xi(p^2)-p^2-\mu^2-\alpha
   {\rm ~e}^{-\frac{\beta^2}{2}\int^\Lambda \xi^{-1}}\right)
\label{min}
\end{equation}
 The equation for the minimum is solved by
\begin{equation}
\xi(p^2)=p^2+M^2
\label{ans}
\end{equation}
where the mass parameter $M^2$ satisfies the equation
\begin{equation}
M^2=\mu^2+\alpha
\left(\frac{M^2/\Lambda^2}{1+M^2/\Lambda^2}\right)^{\beta^2/8\pi}
\label{cond}
\end{equation}
We substitute (\ref{ans}) into (\ref{var}) and integrate to obtain
\begin{equation}
{1\over V}W\leq\frac{1}{8\pi}\left(
\Lambda^2\ln\left(\frac{\Lambda^2+M^2}{\Lambda^2}\right)
+\mu^2\ln\left(\frac{\Lambda^2+M^2}{M^2}\right)-
8\pi\frac{\alpha}{\beta^2}\left(
\frac{M^2}{\Lambda^2+M^2}\right)^{\beta^2/8\pi}
\right)
\label{pot}
\end{equation}
The regime that we are interested in is where the cutoff, (the
electron mass) is very large and the other dimensional parameters
$\mu$ and $\alpha$ are very small.  We look for a minimum of the
potential (\ref{pot}) where $M$ is of the same order of magnitude as
$\mu$ and $\alpha$.  This corresponds to seeking a solution of
sine-Gordon theory which is consistently renormalized as a
relativistic quantum field theory.

\subsubsection{Sine-Gordon theory ($\mu^2=0$, $\Lambda\rightarrow\infty$)}

It is interesting to explore the minima of (\ref{pot}) in the case of
pure sine-Gordon theory, when $\mu^2=0$.  This can be done using a
slightly different variational method by Coleman \cite{co}.

We must seek a solution of the equation for the variational mass where
$M^2<<\Lambda^2$.  To do this, we must first renormalize the bare
coefficient of the cosine term in the action,
\begin{equation}
\alpha\equiv \alpha_R\left( \frac{\Lambda^2}{a^2}\right)^{\beta^2/8\pi}
\end{equation}
where $a$ is an arbitrary mass scale which accompanies
renormalization.  Then, up to an infinite, $M$ independent constant,
the variational free energy is
\begin{equation}
\frac{1}{V}W\leq\frac{1}{8\pi}\left( M^2-8\pi\frac{\alpha_R}{\beta^2}
\left( \frac{M^2}{a^2}\right)^{\beta^2/8\pi}\right)
\end{equation}
Clearly, when $\beta^2/8\pi < 1$, the point $M=0$ is a local maximum
of the potential.  There is a minimum at $M\neq 0$, determined by the
finite dimensional parameters $a$ and $\alpha_R$ and by $\beta$.  On
the other hand, when $\beta^2/8\pi>1$, there is a local minimum of the
potential at $M=0$ but for large $M$, it is unbounded from below, This
means that $M=0$ is an unstable state and the value of $M$ at the true
minimum of the potential is of order the (infinite) cutoff.

Thus, we regain Coleman's result~\cite{co}. Without the renormalization of
the $\beta^2$ parameter, which was shown to be necessary in ref.~\cite{am},
the pure sine-Gordon model has a phase transition at 
the point $\beta^2_c=8\pi$ from a
phase where the theory is approximately solved by a boson with finite
mass to one where the boson has infinite mass, of order of the
cutoff.  

\subsubsection{Sine-Gordon theory: parity invariant QED
($\mu^2=0$, $\Lambda$ finite )}

Let us now analyze the case in which the cutoff is finite and of the order
of the electron mass $m$. We can then analyze directly the potential  
(\ref{pot}) with $\Lambda\sim m$ and $\mu=0$. It can be easily seen that, 
independently on the value of $\Lambda$, the potential (\ref{pot}) for 
$\beta^2/8\pi < 1$ behaves as before, $i.e.$ at the point $M=0$ has a local
maximum and an absolute minimum at $M\neq 0$. For $\beta^2/8\pi > 1$,
however the minimum at $M=0$ is now an absolute minimum for any 
$M<\Lambda$, the potential is always bounded below. 
$M=0$ describes a stable state, so that the transition is between a
phase characterized by the behavior of a two-dimensional massless boson 
to one charcterized by a massive boson.

This can be interpreted as a change in the symmetry, since a
massless boson has the field translation symmetry, $\phi(\vec x)
\rightarrow \phi(\vec x)+{\rm const.}$ whereas a massive boson does not.
If we translate the parameters of the sine-Gordon theory into those of
the effective action for QED (as in the discussion after equation
(\ref{sg})), the phase transition occurs at the critical temperature
$T_c=e^2/8\pi$. Above this temperature, the boson has a mass,
corresponding to deconfined phase which screens electric charges by
virtue of the ``Debye'' mass $M$ (region III in Fig.1), whereas below
this temperature the boson is massless (region I in Fig.1), 
corresponding to a confining phase which cannot
screen the long ranged electric fields of incommensurate charges.

\subsubsection{Sine-Gordon theory with a mass
($\mu^2\neq 0$, $\Lambda
\rightarrow\infty$)}

Let us now consider the case of massive sine-Gordon theory with
$\mu^2>0$.  The variational potential is made finite by the same
renormalization of the parameter $\alpha$ as in the massless case.
The variational free energy  is then
\begin{equation}
\frac{1}{V}W=\frac{1}{8\pi}\left( M^2-\mu^2\ln(M^2/a^2)-8\pi
\frac{\alpha_R}{\beta^2}\left( \frac{ M^2}{a^2}\right)^{\beta^2/8\pi}
\right)
\end{equation}
If $\beta^2<8\pi$, the minimum of the potential occurs at a finite
value of $M^2$. This value depends crucially also on the topological
mass $\mu$ and for $\mu$ very small occurs approximately at the same
value of the $\mu=0$ case, for $\mu$ sufficiently large it occurs at
$\mu$.  As in massless sine-Gordon theory, without a renormalization
of $\beta^2$, the potential is unbounded below if $\beta^2>8\pi$,
independently on the value of $\mu$.  This implies that the global
minimum of the potential is of order the ultraviolet cutoff.

\subsubsection{Sine-Gordon theory with a mass:
topologically massive QED ($\mu^2\neq 0$, $\Lambda$ finite)}

For $\mu$ very small and $\beta^2<8\pi$ (high temperature) the
potential (\ref{pot}) has an absolute minimum at a finite value of $M$
much larger than $\mu$, independently on the value of the cutoff
$\Lambda$ . For $\beta^2>8\pi$ (low temperature) the absolute minimum
rapidly moves to the small value $M\simeq\mu$, so that the transition
is between two distinct massive behaviors. For a very tiny $\mu$ the
crossover in the BKT should arise only at very large scales.  The
behavior of the system in the two phases should be quite different.
In the low temperature region all the components of the gauge field
have a small mass (the topological mass), and this phase should be
very much like the Higgs phase of the Schwinger model~\cite{gsst}.  In
the high temperature region, the electric mass grows to the much bigger
Debye mass. This region is essentially similar to the plasma phase
of the parity invariant theory. Moreover, the Debye screening mass is
temperature dependent, whereas the topological mass does not
depend on temperature.

It is interesting that this transition occurs even when there is an
explicit mass term in the action, provided that the mass term is
sufficently small. Although this transition is in a sense associated
with vortex binding-unbinding, just as it is in the sine-Gordon theory
which describes the Coulomb gas, it is not a confinement-deconfinement
transition in the strict sense, since the $Z$ symmetry is broken in
both phases.

When the topological mass $\mu$ is large, the minimum of the potential
is always at $\mu$ regardless of the value of $\beta^2$, so that the
crossover arises at short scales ($\sim 1/\mu$) and the flow diagram
is completely destroyed. No drastic change in the behavior of the
system should be observed as the temperature is changed in the
neighborough of $T_c=e^2/8\pi$.

\section{Discussion}

In this paper, we have shown that deconfinement in finite temperature
QED can be characterized as breaking of a certain global discrete
symmetry.  We have also shown that a confinement-deconfinement
transition takes place in parity invariant 2+1-dimensional QED, at
least in the regime where the electron mass is large.

In a sense, the latter fact is no surprise.  When the electromagnetic
coupling $e^2$ is small compared to the electron mass $m$ so that
vacuum fluctuations are suppressed, and when the temperature is also
smaller than the electron mass, the thermal state is to a good
approximation a dilute two-dimensional neutral Coulomb gas of thermally
excited electrons and positrons.  It is well known that this Coulomb
gas, even when very dilute, has a BKT transition.  Below the
transition temperature, the electrons and positrons are bound into
pairs.  Above the transition temperature, electrons and positrons are
approximately free particles.  One physical prediction which can be
deduced from the presence of the BKT transition is the universal
property of the phase transition associated with the bulk modulus of
spin waves in the gapless, Kosterlitz-Thouless phase.  The exponents
in the correlators of Polyakov loop operators in that phase are
determined by a single correlation length divided by the temperature.

There is a similar picture of the topologically massive theory.  If
the topological mass is small, at weak coupling and temperatures
somewhat less than the electron mass, the thermal state is to a good
approximation that of the dilute Yukawa gas.  There is a rapid
change between a quasi-confining and a deconfined behavior close to
the BKT critical point of the parity invariant theory.

In both cases, our analysis is only valid where the electron mass is
large.  We expect that the phase transition, or more correctly the BKT
line of phase transitions, persists for some time as we lower the
electron mass or raise the electric charge.  However, the resulting
strong coupling regime is out of the domain of validity of our
analysis.

An interesting, experimentally testable consequence of the breaking of
the $Z$ symmetry is the existence of domain walls.  For the existence
of such, it is enough to have field configurations that are
non-contractable loops in the $U(1)$ group. The question of domain
walls has been addressed in gluodynamics \cite{3.,4.}  and massless
QED \cite{sm}. Whether they correspond to real Minkowski space
objects, is an interesting open question.  It is instructive to note
that upon refermionizing our effective sine-Gordon model (\ref{sineG})
along the lines of \cite{co}, domain walls in sine-Gordon theory
correspond to worldlines of fermions in the resulting Thirring model.

Phase transitions analogous to the one which we conjecture to exist in
topologically massive QED have been studied experimentally in quasi
two-dimensional condensed matter systems, particularly charged vortex
arrays in superconducting films ~\cite{exp} and have also been useful
in theoretical work on high $T_C$ superconductivity ~\cite{sta} where
a type of ``smoothed'' BKT transition is discussed.  The experimental
study of analog systems in condensed matter physics could help to
resolve some of the theoretical questions raised by our current work.

\section*{Acknowledgement}

We thank I. Affleck, A. D'Adda, R.  Ferrari, A. Kovner, I.  Kogan,
A. Polyakov, P. C. E. Stamp, L.C.R. Wijewardhana and N.  Weiss for
helpful discussions.  G. Semenoff and O. Tirkkonen acknowledge the
kind hospitality of the Dipartimento di Fisica, Universit\`a di
Perugia and Universit\`a di Trento where part of this work was done.

\pagebreak

\null

\vspace{-7cm}

\hspace*{-40truemm}
\epsffile{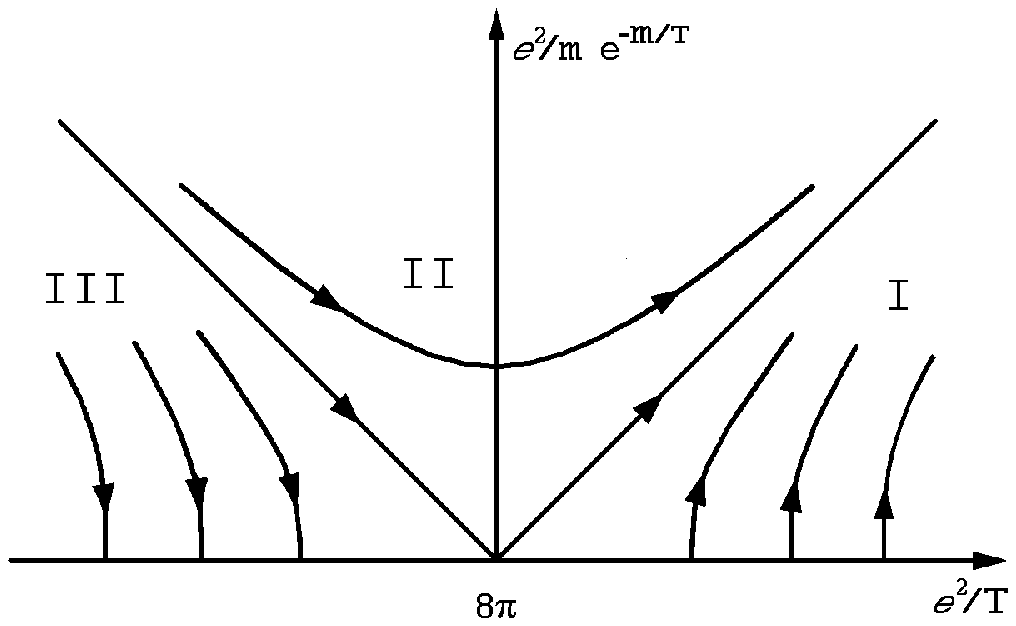}

\vspace{-9.4truecm}

\centerline{Figure 1.}

\bigskip

\noindent
Renormalization flow diagram for the BKT transition.
The arrows denote flow toward the ultraviolet.  Region I is the confining
phase whereas regions II and III are deconfined.  Region III is
asymptotically free whereas in region I there is a line of infrared
stable fixed points which represent $c=1$ conformal field theories.
The separatrix between regions I and II is the line of BKT phase
transitions.  The critical behaviour of the system at the latter phase
transition is that of an SU(2) Wess-Zumino-Witten model.

\end{document}